\documentclass[aps,11pt,nofootinbib]{revtex4-1}
 \pdfoutput=1
\usepackage{graphicx}
\usepackage[margin=1in]{geometry}
\usepackage{amsmath}
\usepackage{amssymb}
\usepackage{color}
\usepackage[utf8]{inputenc}
\usepackage[normalem]{ulem}
\usepackage{siunitx}
\usepackage{booktabs}
\usepackage{tabu}

\makeatletter
\renewcommand\@make@capt@title[2]{%
  \@ifx@empty\float@link{\@firstofone}{\expandafter\href\expandafter{\float@link}}%
   {\textbf{#1}}\@caption@fignum@sep#2\quad
}%
\makeatother

\begin{document}

\author{Sam Young}
\email{syoung@mpa-garching.mpg.de}

\affiliation{Max Planck Institute for Astrophysics, Karl-Schwarzschild-Strasse 1, 85748 Garching bei Muenchen, Germany}

\date{\today}

\title{The primordial black hole formation criterion re-examined: parameterisation, timing, and the choice of window function}

\begin{abstract}
In this paper, the criterion used to determine whether a density perturbation will collapse to form a primordial black hole (PBH) is re-examined, in respect of its use to determine the abundance of PBHs. There is particular focus on which parameter to use, the time at which the abundance should be calculated, and the use of different smoothing functions. It is concluded that, with the tools currently available, the smoothed density contrast should be used rather than the peak value, and should be calculated from the time-independent component of the density contrast in the super-horizon limit (long before perturbations enter the horizon) rather than at horizon crossing. For the first time the effect of the choice of smoothing function upon the formation criterion is calculated, and, for a given abundance of PBHs, it is found that the uncertainty in the amplitude of the power spectrum due to this is $\mathcal{O}(10\%)$, an order of magnitude smaller than previous calculations suggest. The relation between the formation criterion stated in terms of the density contrast and the curvature perturbation $\mathcal{R}$ is also discussed.
\end{abstract}

\maketitle

\tableofcontents

\section{Introduction}

Primordial black holes (PBHs) are black holes which may have formed in the early universe. In order for a perturbation to collapse to form a PBH, the density must exceed some critical threshold. The original work by Carr \cite{Carr:1974nx} provided an order of magnitude estimate that the density contrast should be above one third at horizon crossing in order for a PBH to form, $\delta_c>1/3$. In recent years, there has been a great deal of work to determine the collapse threshold \cite{Musco:2004ak,Musco:2008hv,Musco:2012au,Musco:2018rwt,Harada:2015ewt,Harada:2015yda,Nakama:2013ica,Nakama:2014fra,Shibata:1999zs,Niemeyer:1999ak,Polnarev:2006aa}, as well as discussion about the appropriate parameter to use to determine whether a perturbation will collapse \cite{Green:2004wb,Young:2014ana}. Normally, the collapse threshold is calculated from numeric simulations, although analytic calculations neglecting pressure gradients have been made \cite{Harada:2013epa}.

As a parameter to determine PBH formation, the density contrast $\delta$ should be used rather than a metric perturbation such as the curvature perturbation $\mathcal{R}$. Whilst there has been valid work to determine a collapse threshold in terms of a metric perturbation (i.e \cite{Shibata:1999zs}), these are typically only valid when a single isolated perturbation (in a flat background universe) is considered - which this is unlikely to be the case, and environmental effects are likely to change the collapse threshold significantly \cite{Young:2014ana,Harada:2015yda}.

There has been much ambiguity in the literature about how this critical amplitude is calculated and used (especially between the different communities of relativists modelling PBH formation and cosmologists calculating the abundance of PBHs), and it is the aim of this paper to clarify how this should be defined and utilised to make calculations of the PBH abundance. PBHs forming from the collapse of large density perturbations upon horizon re-entry will be considered here, the mechanism for which is described briefly below:

As the comoving Hubble horizon shrinks during inflation, quantum fluctuations become classical density perturbations as they exit the horizon. Once inflation ends, the comoving Hubble horizon begins to grow again during the radiation dominated epoch of the early universe. Once a perturbation re-enters the horizon, if the amplitude of the perturbation is large enough and gravity outweighs the pressure forces, it will collapse to form a PBH. Otherwise, if the perturbation is not dense enough and gravity not sufficiently strong, the perturbation is quickly damped out. 

The density contrast $\delta(x,t)$ will be defined in the comoving, uniform-cosmic time gauge as
\begin{equation}
\delta(x,t)= \frac{\rho-\rho_b}{\rho_b} 
\end{equation}
where $\rho$ and $\rho_b$ are the density and background density respectively. At the linear level on super-horizon scales, the density contrast is related to the curvature perturbation $\mathcal{R}$ as
\begin{equation}
\delta(x,t) = -\frac{2(1+\omega)}{5+3\omega} \left(\frac{1}{aH}\right)^{2} \nabla^2\mathcal{R}(x),
\end{equation}
where $\omega=1/3$ is the equation of state during radiation domination, $(aH)^{-1}$ is the Hubble horizon (with $a$ and $H$ being the scale factor and Hubble parameter respectively), and $\nabla^2$ is the Laplacian. This can be expressed in Fourier space as
\begin{equation}
\delta(k,t) = -\frac{2(1+\omega)}{5+3\omega} \left(\frac{k}{aH}\right)^{2} \mathcal{R}(k).
\end{equation}
Since $\mathcal{R}$ is conserved on super-horizon scales for adiabatic perturbations and the density contrast on super-horizon scales grows proportionally to the horizon scale squared, the density contrast can be separated into time-dependent and time-independent components for a given perturbation of scale $r_m$,
\begin{equation}
\delta(x,t)  = \epsilon^2(t) \delta_{\mathrm{TI}}(x),
\label{eqn:SHdelta}
\end{equation}
$\delta_{\mathrm{TI}}(x)$ is the time-independent component of the density contrast. The parameter $\epsilon(t)$ is the ratio of the horizon scale at some time $t$, $(aH)^{-1}$, and the scale of the perturbation $r_m$,
\begin{equation}
\epsilon(t)=(aHr_m)^{-1}.
\end{equation}


Throughout the paper spherical symmetry will be assumed, which is typical when studying PBHs, justified by the fact that peaks which form PBHs are large and rare \cite{Bardeen:1985tr} - although non-spherical symmetry has been considered \cite{Harada:2015ewt}. For simplicity, it will also be taken that the PBH formation process occurs during radiation domination, although phase transitions which affect the equation of state of the universe have a significant impact on the abundance of PBHs \cite{Byrnes:2018clq}.

The paper is organised as follows: in section \ref{sec:shapes} the typical profile shapes of primordial perturbations will be discussed, in section \ref{sec:musco} the threshold value derived from simulations is discussed, section \ref{sec:time} discusses the time at which PBH abundance should be calculated and the time at which PBHs form, section \ref{sec:smooth} discusses using the peak value of the density or the volume-averaged value as the formation criterion, section \ref{sec:smoothingfunction} discusses the calculation of PBH abundance using different window functions, \ref{sec:zeta} discusses the formation criteria as calculated from the curvature perturbation $\mathcal{R}$, and the conclusions of the paper are summarised in section \ref{sec:summary}.

\section{Average profile shapes}
\label{sec:shapes}
The shape of primordial density fluctuations is unknown, and is likely to depend upon the specific model of inflation being considered. However, the average profile shape close to the centre of a perturbation can be estimated from the power spectrum $\mathcal{P}_\delta$ (see \cite{Germani:2018jgr} for further discussion). Assuming spherical symmetry, the profile shape can be conpletely described in terms of the radial co-ordinate $r$, the distance from the centre of the peak,
\begin{equation}
\delta(r) = \delta(0)\frac{\xi(r,t)}{\xi(0,t)},
\label{eqn:deltaShapes}
\end{equation}
where $\delta(0)$ is the amplitude at the centre of the perturbation. $\xi(r,t)$ is given by
\begin{equation}
\xi(r,t) = \frac{1}{(2 \pi)^3}\int \frac{\mathrm{d}k}{k}\frac{\mathrm{sin}(kr)}{kr}\mathcal{P}_\delta(k,t),
\end{equation}
where this formalism requires the power spectrum to converge to zero in the UV and IR limit (a ``cut-off") in order to predict a profile shape, which happens naturally in the IR limit due to the $k^4$ factor in the power spectrum. Due to the $k^2$ super-horizon growth of perturbations, the derived average profile shape is typically dominated by the largest values of $k$ before the UV cut-off. 

\begin{figure*}
 \centering
  \includegraphics[width=0.49\textwidth]{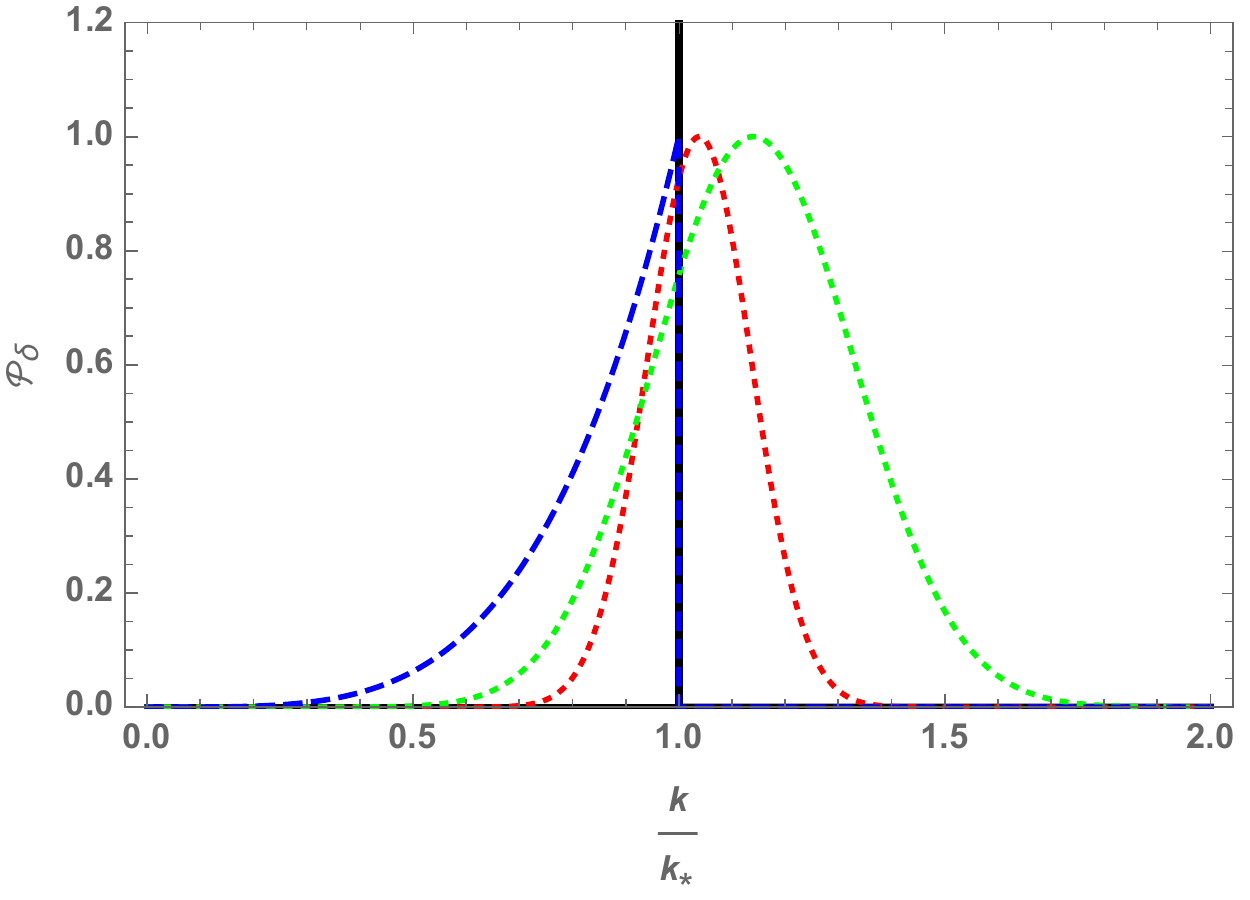} 
  \includegraphics[width=0.49\textwidth]{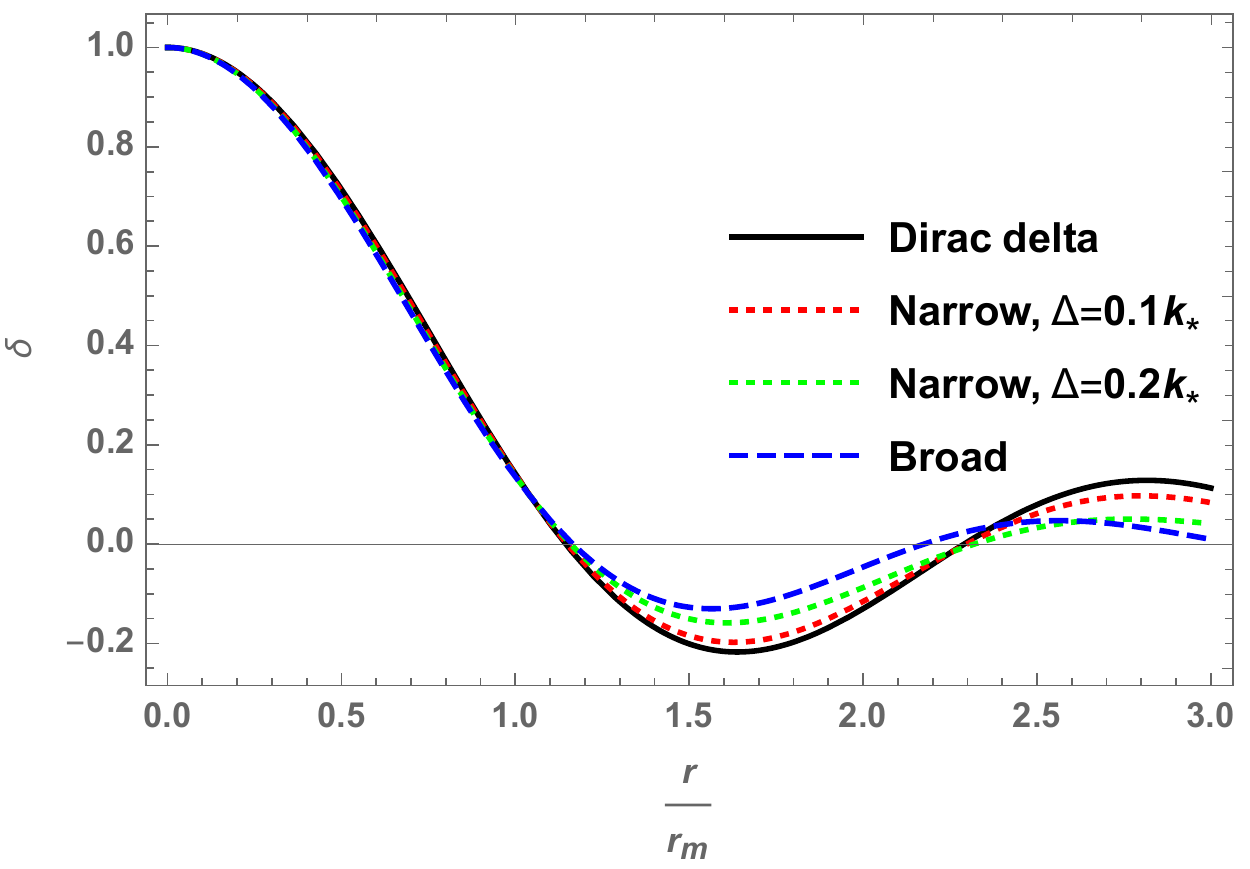} 
  \caption{ The left plot shows some of the shapes of the primordial power spectrum often considered in PBH literature. The shapes are as follows: an (unphysical) Dirac delta shape (black, solid); a narrow peak in the power spectrum (red and green dotted for $\Delta=0.1k_*,0.2k_*$ respectively), and a broad power spectrum with a sharp $k$ cut-off (blue, dashed). The average profile shape of perturbations for each power spectrum shape is shown on the right. The profiles have been normalised to have the same central amplitude and characteristic scale $r_m$.  }
  \label{fig:shapes}
\end{figure*}

Several power spectra often considered in the literature are considered here:
\begin{itemize}
\item the \emph{Dirac-delta} power spectrum (although it is noted that this not physical \cite{Byrnes:2018txb})
\begin{equation}
\mathcal{P}_\delta (k,t)= \mathcal{A} \left(\frac{k}{aH}\right)^4 \delta_D(k-k_*).
\end{equation}
\item the \emph{narrow-peak} power spectrum, typically described with a Gaussian peak,
\begin{equation}
\mathcal{P}_\delta(k,t) = \mathcal{A} \left(\frac{k}{aH}\right)^4 \exp\left(- \frac{(k-k_*)^2}{2 \Delta^2} \right),
\end{equation}
where $\Delta$ paramterises the width of the peak. Also often considered is a log-normal peak,
\begin{equation}
\mathcal{P}_\delta(k,t) = \mathcal{A} \left(\frac{k}{aH}\right)^4 \exp\left( -\frac{\mathrm{log}(k/k_*)^2}{2 \Delta^2} \right).
\end{equation}
For narrow peaks, $\Delta \ll 1$, this is similar to the Gaussian peak, and will therefore not be considered further.
\item and the \emph{broad-peak} power spectrum (which is scale-invariant over the width of the peak)
\begin{equation}
\mathcal{P}_\delta(k,t) = \mathcal{A} \left(\frac{k}{aH}\right)^4 \Theta(k-k_{min})\Theta(k_*-k),
\end{equation}
where $\Theta(k)$ is the Heaviside step function, and $k_{min} \ll k_*$.
\end{itemize}
In each case, the amplitude of the power spectrum is parameterised by $\mathcal{A}$, the $(k/(aH))^4$ term represents the super-horizon growth of perturbations, and the power spectra drop quickly to zero for $k>k_*$. All of the density power spectra can be considered as qualitatively similar - being well represented by a narrow peak at or close to $k_*$, quickly dropping to zero either side of the peak. The left plot of figure \ref{fig:shapes} shows the power spectra described, normalised to have a maximum amplitude of unity (except for the Dirac delta function).

The characteristic scale of a perturbation is best described using the value $r_m$, which is the radius at which the compaction function $\mathcal{C}(r)$ is maximized
\begin{equation}
\mathcal{C}(r) = 2\frac{M(r,t)-M_b(r,t)}{R(r,t)},
\label{eqn:comp}
\end{equation}
where $R(r,t)$ is the areal radius, $M(r,t)$ is the mass within a sphere of the radius centred on the peak and $M_b(r, t)$ is the background mass within the same areal radius calculated with respect to a flat FRW Universe. The characteristic scale $r_m$ of the profiles varies by a factor of $\mathcal{O}(1)$ for the power spectra considered here, and as mentioned earlier, the scale $r_m$ of each perturbation is of the same order as $1/k_*$. The right hand plot of figure \ref{fig:shapes} shows the average profile shapes for the difference profile shapes, normalised to have an amplitude equal to unity in the centre and plotted with respect to their respective characteristic scales. The average profiles are extremely similar, especially in the central region close to the peak. For the remainder of this paper, it will therefore be taken that the representative profile shape is given by that of the Dirac delta power spectrum, which has a simple analytic form,
\begin{equation}
\delta(r) = A \frac{\mathrm{sin}(2.744 r)}{2.744 r},
\label{eqn:sinc}
\end{equation}
where the amplitude is parameterised by $A$, and the numerical factor or $2.744$ is included for simplicity such that $r_m=1$ (in arbitrary units). This equation does not account for the time-evolution of $\delta$, and can be considered as the time-independent component of equation \eqref{eqn:SHdelta}. 

The formalism applied in this section is utilised only to determine a `standard' profile shape (not used in determining the abundance), and will be used in section \ref{sec:musco} to determine a single critical value for collapse - which in principle takes a range of values depending on the exact profile shape. For a broad power spectrum, large perturbations will exist on a range of scales. However, as seen, this procedure results in perturbations which all have a characteristic scale $r_m \sim 1/k_*$ - and does not predict the shapes of perturbations on larger scales. A smoothing function of a given scale $R$ can first be applied to the power spectrum, and then the average profile shape calculated - which gives average perturbations on a scale similar to $R$, with a very similar profile shape to equation \eqref{eqn:sinc}. However, this means that the smoothing function used has a physical effect on the profile shape, and since the profile shape affects the evolution of the perturbation, this leads to the false conclusion that the smoothing function has a physical effect on the evolution of that patch of the universe. Rather than apply this therefore, it will simply be taken that equation \eqref{eqn:sinc} is still representative of such perturbations. An investigation into the probable shape of such perturbations is beyond the scope of this paper.

\section{The Musco criterion from simulations}
\label{sec:musco}

A great deal of work has been completed in recent years to calculate the threshold value for a PBH to form, involving the use of numerical and analytic methods \cite{Musco:2004ak,Musco:2008hv,Musco:2012au,Musco:2018rwt,Harada:2015ewt,Harada:2015yda,Nakama:2013ica,Harada:2013epa,Nakama:2014fra}. Typically, an initial density perturbation is defined on super-horizon scales, and evolved forwards through horizon re-entry to determine whether a PBH will form from the perturbation or else dissipate into the surrounding universe. 

The most commonly used parameter used to describe the amplitude of a perturbation is the volume averaged density contrast $\delta_R$ (although the central height of a peak $\delta_{pk}$ is also used, discussed later in section \ref{sec:smooth}). In order to calculate $\delta_R$ the density contrast is integrated over its characteristic volume (a sphere of radius $r_m$) and divided by the volume:
\begin{equation}
\delta_R = \frac{1}{V}\int\limits_0^{r_m}\mathrm{d}r 4 \pi r^2 \delta(r,t),
\end{equation}
where $V=4\pi r_m^3/3$. The time at which the amplitude is evaluated will be discussed further in section \ref{sec:time}, although the amplitude is typically stated by Musco in terms of the time-independent component of the density perturbation. See \cite{Musco:2018rwt} for a more detailed discussion. A number of simulations are run in order to determine the critical amplitude of a perturbation in order for a PBH to collapse, $\delta_c=\delta_{R,critical}$ - which will be referred to as the Musco criterion for the rest of this paper, and is widely used as the standard formation criterion. This value is found to vary significantly depending on the profile shape, $0.41<\delta_c<2/3$. For the representative profile shape considered here (given in equation \eqref{eqn:sinc}), the critical amplitude is $\delta_c\approx 0.51$ (or $\delta_{c,pk}\approx1.2$ if using the peak value) \cite{Germani:2018jgr}.

\section{Timing is everything}
\label{sec:time}

In this section, the time at which perturbations should be evaluated to determine whether they will form a PBH, as well as the time at which PBHs can be considered to form will be discussed.

\subsection{When should you calculate PBH abundance?}
\label{sec:calctime}

\begin{figure*}
 \centering
  \includegraphics[width=0.8\textwidth]{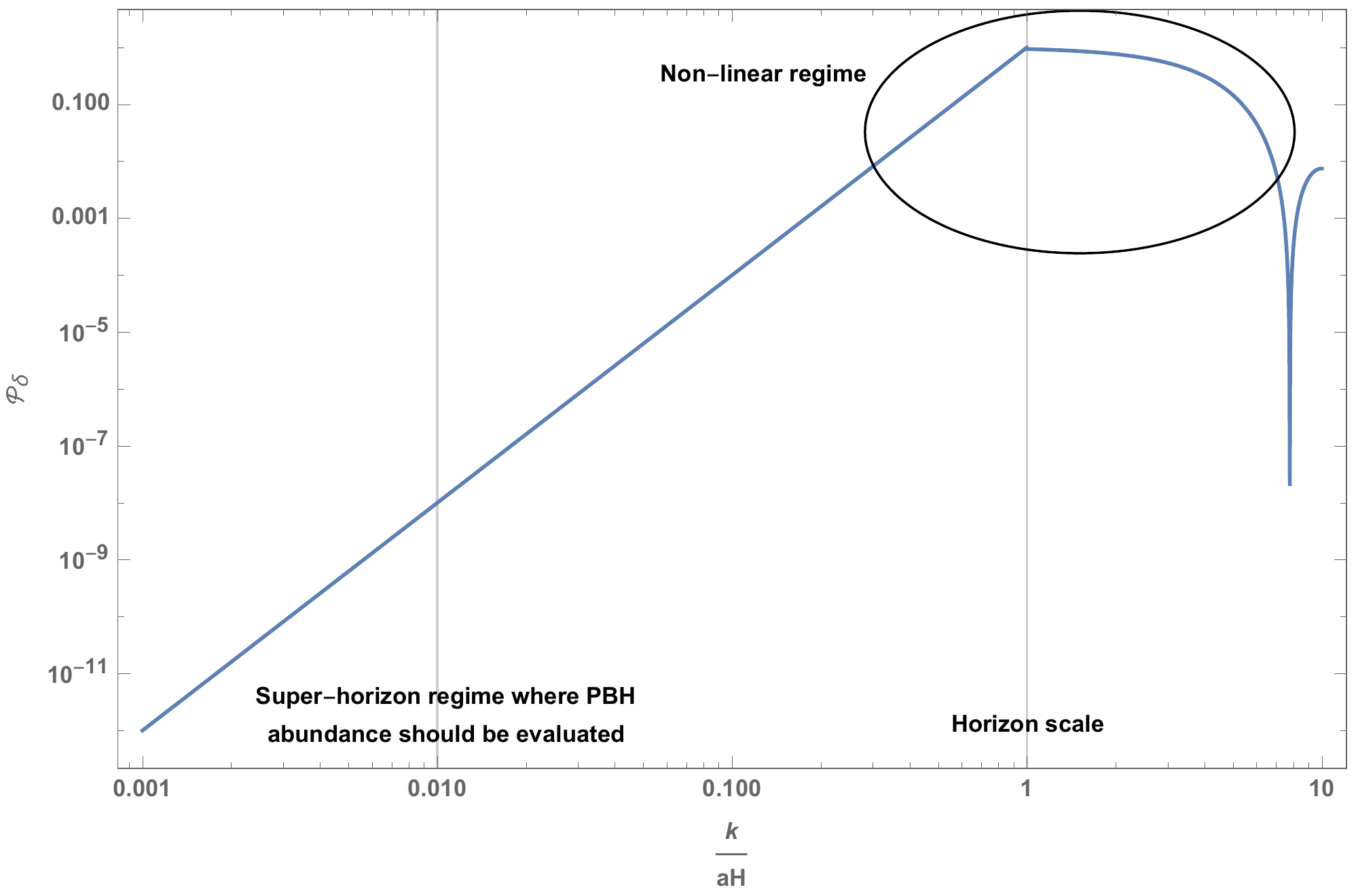} 
  \caption{ A schematic diagram of the evolution of the primordial power spectrum during horizon re-entry, relevant for PBH formation. On super-horizon scales, the power spectrum grows proportionally to $(k/(aH))^4$, becoming large and non-linear around the time of horizon re-entry. After horizon re-entry, the power-spectrum shows oscillatory behaviour as pressure- and gravity-forces smooth out perturbations.}
  \label{fig:power}
\end{figure*}

Figure \ref{fig:power} shows a schematic the power spectrum $\mathcal{P}_\delta$ as a function of $k/(aH)$. The graph can be interpreted in 2 ways: it can be considered to show the time-evolution of the power spectrum at a given scale as the horizon scale grows, or it can be considered that the (scale invariant) power spectrum at different scales $k$ is being considered at a given time, corresponding to a given horizon scale $(aH)^{-1}$.

In the first picture, it is seen that the power spectrum grows rapidly in the super-horizon regime as they near the horizon scale, but is then quickly damped out by radiation forces after horizon entry. A PBH will form (or not) when a given perturbation enters the horizon, and so the natural conclusion is that only modes with a scale similar to that of the perturbation scale will contribute significantly to PBH formation. However, close to horizon crossing the gradient expansion approach (which is used in the derivation of equation \eqref{eqn:nonlinear} seen later) breaks down, and perturbations forming PBHs are necessarily large and non-linear - requiring the use of complicated simulations to model them. In order to accurately calculate the abundance of (very rare) PBHs in the (very large) universe, it is therefore desirable to be able to define whether or not a PBH will form by evaluating the linear initial conditions of a perturbation. It will be argued here that the specific time that this is done is unimportant, provided that perturbations are evaluated in the super-horizon regime, and that the time-independent component should be used. 

Considering now the second scenario (that figure \ref{fig:power} is showing the amplitude of an initially scale invariant power spectrum at a given time) it is seen that the distribution of matter in the Universe will be dominated by the large amplitude modes near horizon crossing. This therefore motivates the use of a smoothing function to remove these modes artificially from the calculation so that larger scales can be considered, which will be discussed further in section \ref{sec:smoothingfunction}. For the remainder of this section, it will be assumed that at the time considered, the power spectrum has some UV cut-off on super horizon scales \footnote{This may be an artificial cut-off from a smoothing function or a natural cut-off in the primordial power spectrum} such that the entire distribution can be treated as super-horizon.

The relative height of a peak is often described in both peaks theory and Press-Schechter using the parameter $\nu$, which is given by \begin{equation}
\nu = \frac{\delta}{\langle \delta^2 \rangle ^{1/2}},
\label{eqn:nu}
\end{equation}
The relative height $\nu$ (more specifically, the critical value $\nu_c=\delta_c/\langle \delta^2 \rangle ^{1/2}$) is the crucial parameter for determining PBH abundance. The full calculation will not be discussed here, and can be significantly affected by primordial non-Gaussianity \cite{Young:2013oia,Young:2014oea}, and the non-linear relations between the density contrast and the curvature perturbation (discussed briefly in section \ref{sec:zeta}, readers are directed to recent papers \cite{Young:2019yug,Yoo:2018esr,Kawasaki:2019mbl,DeLuca:2019qsy}).
In equation \eqref{eqn:nu}, $\langle \delta^2 \rangle$ is the variance of perturbations, which can be found by integrating over the power spectrum:
\begin{equation}
\langle \delta^2 \rangle  = \int\limits_0^\infty \frac{\mathrm{d}k}{k}\mathcal{P}_\delta.
\label{eqn:powerintegral}
\end{equation}
For simplicity, smoothing and transfer functions are neglected here, and it will be taken that these have already been applied to the power spectrum. It is noteworthy at this point that $\mathcal{P}_\delta$ must have a UV cut-off in order for this integral to converge (the $k^4$ term naturally gives an IR cut-off). At the linear level, on super-horizon scales, the density power spectrum $\mathcal{P}_\delta$ can be related to the curvature perturbation power spectrum $\mathcal{P}_\mathcal{R}$ during radiation domination as
\begin{equation}
\mathcal{P}_\delta = \frac{16}{81}\left(\frac{k}{aH}\right)^4\mathcal{P}_\mathcal{R} = \frac{16}{81}\left(\frac{1}{aHr_m}\right)^4\left( kr_m \right)^4\mathcal{P}_\mathcal{R}= \frac{16}{81}\epsilon^4(t)\left( kr_m \right)^4\mathcal{P}_\mathcal{R}.
\label{eqn:powers}
\end{equation}
The curvature perturbation is known to be conserved on super-horizon scales, assuming adiabatic perturbations \cite{Lyth:2004gb}, and so the complete time-dependence of $\mathcal{P}_\delta$ is contained, again, by the $\epsilon$ parameter, and $\langle \delta^2 \rangle ^{1/2}$ can be separated into into a time-dependent and time-independent component:
\begin{equation}
\langle \delta^2 \rangle = \epsilon^4(t) \int\limits_0^\infty \frac{\mathrm{d}k}{k}\frac{16}{81}\left( kr_m \right)^4\mathcal{P}_\mathcal{R}=\epsilon^4(t)\langle \delta_{TI}^2 \rangle,
\end{equation}
where it is noted that this gives the time-independent component normalised to a scale $r_m$. Therefore, both $\delta$ and $\langle \delta^2 \rangle ^{1/2}$ grow proportionally to $\epsilon^2(t)$ and $\nu$ is also conserved on super-horizon scales,
\begin{equation}
\nu = \frac{\epsilon^2(t) \delta_{TI}}{\epsilon^2(t)\langle \delta^2 \rangle ^{1/2}_{TI}}= \frac{\delta_{TI}}{\langle \delta_{TI}^2 \rangle ^{1/2}},
\end{equation}
justifying the earlier statement that the specific time of evaluation is unimportant and that the time-independent component should be used. This is consistent with using the Musco criterion - which uses the time-independent component of the initial density perturbation to state the critical amplitude.

Because using the time-independent component can be considered mathematically equivalent to setting $\epsilon=1$, this is often considered to be the same thing as evaluating perturbations at horizon entry - which is not strictly correct as the more complicated behaviour at horizon entry is neglected. The distinction here may seem trivial, but becomes important when one considers that linear transfer functions are often applied in the non-linear regime close to horizon entry (this also introduces a time-dependent component ignored in the super-horizon limit, where the transfer function $T(k,t) = 1$). The importance of this distinction will require the use of numerical simulations, and is left for future study.

\subsection{At what time do PBHs form?}

In section \ref{sec:calctime}, the time at which the abundance of PBHs should be calculated was discussed, concluding that this calculation is best performed in the super-horizon regime before the calculations break down close to horizon entry. By this it is meant that that the calculation of their abundance should be calculated from their initial conditions before any PBHs form when $\epsilon \ll 1$, but will argue that their initial abundance is best described at the time when $\epsilon =1$.

Due to the relative red-shift of the PBH density compared to radiation density, the energy fraction of the universe contained within PBHs changes as a function of time following their formation. The abundance of PBHs is typically stated in terms of $\beta$, the energy fraction of the universe contained within PBHs at the time of their formation\footnote{Another commonly used parameter is the present-day fraction of dark matter contained within PBHs, $f$.}, which leads to the consideration of the time at which they form - and therefore at which time their initial abundance should be stated.

Typically, the scale factor increases by a factor $\mathcal{O}(3)$ between horizon crossing and PBH formation \cite{Musco:2018rwt}, but can take significantly longer for perturbations extremely close to the critical value \cite{Musco:2008hv}, and horizon crossing can happen at slightly different times due to the effect of super-horizon modes - meaning that an exact time cannot be given. However, a time that they are taken to form can be given.
Strictly speaking, this is a free choice as it is not possible to observe them at the time of formation, only their effects at some later time - meaning that as long as you calculate consistently the abundance at the later time, the exact initial time is not important (see the next paragraph). What is important is to reach a consensus so that results and constraints from different sources can be applied consistently.

To consider a concrete example, assume that PBHs form from perturbations crossing the horizon at a time when the horizon mass is equal to $1M_\odot$, and one wishes to know the abundance of PBHs at matter-radiation equality, where the horizon mass will be taken as $2.8\times 10^{17} M_\odot$ \cite{Nakama:2016gzw}. During radiation domination, because of the redshift of energy densities, the density parameter of PBHs $\Omega_{pbh}$ grows proportionally to the scale factor $a$, $\Omega_{pbh} \propto a \propto M_H^{1/2}$, where $M_h$ is the horizon mass. Taking that the formation time of PBHs is the moment of horizon crossing, giving an initial abundance of $\Omega_{pbh}=\beta_{hc}$. At the time of of matter-radiation equality, the density parameter will have grown by a factor $(2.8\times10^{17})^{1/2}$ (assuming the universe to behave exactly as radiation-domination up to this time). Now lets consider that the time of formation is taken to be when the horizon scale has increased by factor of 3 after the perturbations enter the horizon, the initial density parameter is now $\Omega_{pbh}=\beta_{3}=3^{1/2}\beta_{hc}$. Since we are now considering that PBH formation occured at a slightly later time, the density parameter grows only by a factor of $(2.8\times10^{17}/3)^{1/2}$ - giving exactly the same abundance of PBHs evaluated at a later time.

Historically, the time of formation has been considered to be horizon crossing time of the perturbations, and it is only recently in several papers that the fact that they do not form instantly has been taken into account when calculating $\beta$ (i.e. \cite{Germani:2018jgr,DeLuca:2019qsy}). Since the exact time a PBH forms (from a perturbation of a particular scale) is ambiguous and depends upon the amplitude of the individual perturbation, and for the sake of simplicity and convenience, it is recommended here that the time of PBH formation from perturbations of a given scale should be taken as the time that the horizon scale of the background universe is equal to the scale of the perturbation.

\section{To smooth, or not to smooth, that is the question}
\label{sec:smooth}

The use of the volume-averaged (smoothed) density contrast will be compared to using the peak value at the centre of the perturbation. Typically, the abundance of PBHs is calculated using either a peaks theory or a Press-Schechter approach (see \cite{Young:2014ana} for a comparison of the methods), which rely on comparing the height of a peak or average density of a region to a critical value respectively. However, it is often omitted that, when using peaks theory, the peaks are typically those of the smoothed density field and that therefore, the height of a peak corresponds to the average overdensity within the volume of a smoothing function placed at the peak. In this regard then, there is no difference in the formation criteria that should be applied in each case.

A recent paper \cite{Germani:2018jgr} used instead the height of the unsmoothed density contrast as the formation criterion. The critical value was determined by first predicting the average profile shape which would arise from a given power spectrum, and then using numerical results to derive the critical height of such peaks. For each power spectrum considered, either a broad or narrow peak, there was a natural cut-off (without needing a smoothing function) in the power spectrum as $k$ becomes large - meaning smaller-scale modes could be ignored. The effect of larger scales on the profile shape can also be neglected in this picture due to the factor of $k^4$ in the density power spectrum in the super-horizon regime. Such an approach does work, although it can only be used to investigate the formation of PBHs on a single scale. If, however, the power spectrum remained large enough for significant PBH formation on a range of scales, the method proposed in reference \cite{Germani:2018jgr} would not correctly predict the total number of PBHs forming.

\begin{figure*}
 \centering
  \includegraphics[width=\textwidth]{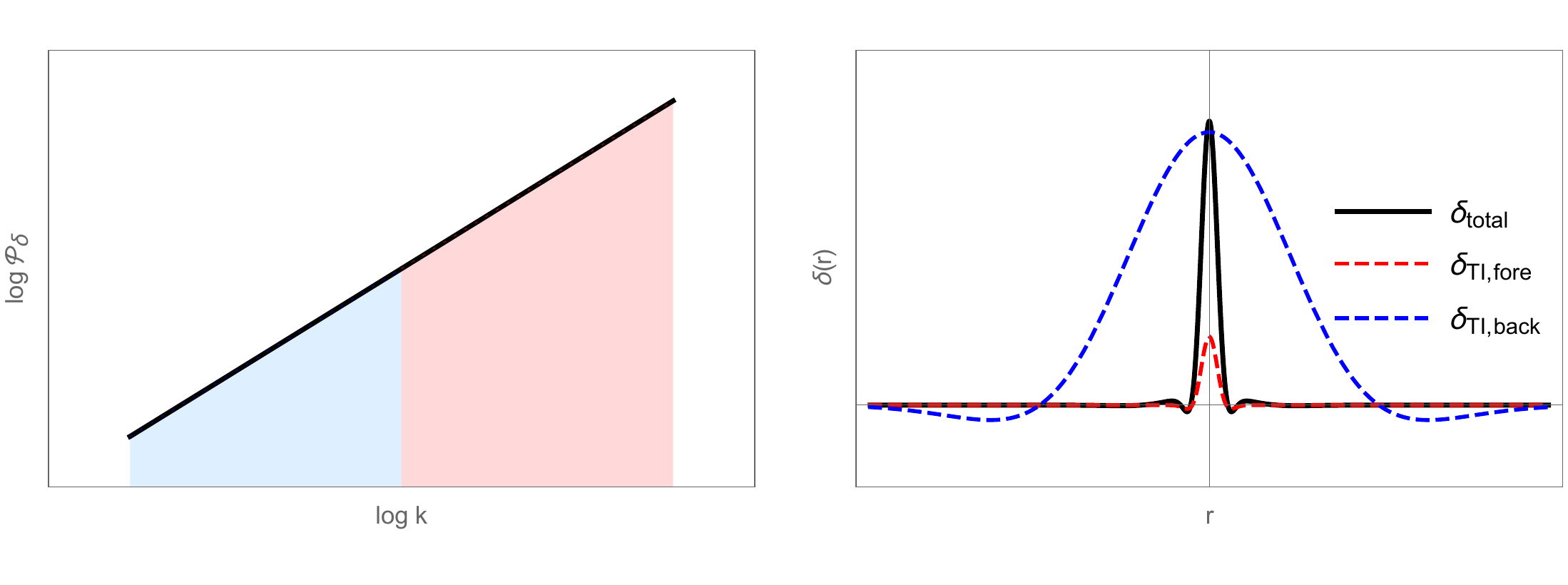} 
  \caption{ The left plot shows a broad peak in the primordial power spectrum, separated into foreground (red) and background (blue) components. The right plot shows a particular realization of the density profile which might be expected from such a power spectrum. The red (blue) dashed line shows the super-horizon time-independent component of the foreground (background) density perturbation. The background perturbation has a much larger amplitude and will collapse to form a PBH upon horizon re-entry, the foreground perturbation will not. However, by looking only at the total density perturbation on super-horizon scales, it is dominated by the foreground perturbation - leading to the conclusion that no PBH forms if only the peak value is considered. }
  \label{fig:foreback}
\end{figure*}

The problem faced is related to the cloud-in-cloud problem faced when calculating galaxy abundances, and will be illustrated by considering a broad peak in the power spectrum, as shown schematically in the left plot of figure \ref{fig:foreback}. The density power spectrum plotted is clearly blue-tilted, but note that the curvature perturbation power spectrum corresponding to this is scale invariant over the width of the peak. Thus, modes of all scales will typically have the same amplitude at horizon-entry, and so PBH formation is equally likely at all scales within the peak.

The average profile shape expected from such a power spectrum will be similar to those shown in figure  \ref{fig:shapes}. However, we will now split the power spectrum into ``foreground'' and ``background'' components, indicated by red and blue respectively in the right plot of figure \ref{fig:foreback}. The foreground component generates a small-scale perturbation, whilst the background generates a large-scale perturbation. The method used to do this is unimportant, and is used simply for demonstration purposes - the important factor is that perturbations of significantly different scales are formed.

For the purposes of this demonstration, in the right plot of figure \ref{fig:foreback} the foreground and background perturbations shown are a factor of 10 different in scale, and the amplitude of the time-independent component of the foreground peak is a factor of 4 smaller than the time-independent component of the background peak - shown by the red and blue dotted lines in the right hand plot respectively (these may also be considered as approximations to the amplitude of each perturbation as it enters the horizon). In this scenario, we can take that the foreground perturbations will not form a PBH upon horizon entry, but the ``larger'' background perturbation will. This may at first seem an unlikely scenario - but is in fact no more unlikely than the formation of PBHs themselves (which are extremely rare events). Consider that we choose a position where a PBH will form from a large-scale perturbation, and then consider the smaller-scale perturbations in this region. It is likely that the small-scale perturbations will not be large enough to form a PBH. However, assuming that perturbations will be studied in the super-horizon limit, and because of the $k^4$ growth of the power spectrum, the small-scale density perturbations will be likely to have a much larger amplitude.

At any given time (in the super-horizon limit), the black line shows the total density perturbation (to arbitrary scaling) - from which it can be seen that the density is dominated by the small-scale foreground perturbation. Without the use of a window function, it is therefore only possible to investigate perturbations of a single scale. As mentioned before, this is similar to the cloud-in-cloud problem, except that in this scenario, the larger cloud also has a very small amplitude\footnote{Strictly speaking, in order to determine on which scales PBHs form in a given region, the correct formalism would be to first to determine whether the largest scale perturbations collapse, and then to investigate smaller and smaller modes until the largest scale at which a PBH forms - as in the excursion set formalism applied to large scale structure of the universe. However, because PBH forming perturbations are so rare, the probability of a smaller PBH forming and then being ``swallowed'' by the formation of a larger PBH at a later time is very small, and has a negligible effect on the abundance.}.

By considering only the peak heights of the density contrast without smoothing, the distribution is dominated by the smallest scale perturbations. However, by smoothing the density on different scales it is possible to investigate perturbations of different scales - and one can correctly identify that the time-independent component of the background perturbation is 4 times greater than that of the foreground perturbation.

 In order to discuss the time dependence, a top-hat smoothing function (corresponding to the volume-averaging in the Musco criterion) will be considered here briefly:
\begin{equation}
W(r,r_m) =\frac{1}{V} \Theta(r_m-r),
\end{equation}
where $\Theta(r)$ is the Heaviside step function, $r_m$ is the smoothing scale, and $V=4\pi r_m^3/3$ is the volume of the smoothing function. Taking that the smoothing function picks out perturbations with a scale $r_m$, and accounting for the fact that this is applied to the time-dependent density contrast, the smoothing function should be multiplied by a factor $\epsilon^{-2}=(aHr_m)^2$ to calculate the time-independent component, we then have:
\begin{equation}
\delta_R = \frac{3}{r_m}(aH)^2\int\limits_0^{r_m}\mathrm{d}r r^2 \delta(r,t),
\end{equation}
which now closely resembles the compaction function, equation \eqref{eqn:comp}, where the integral over the density contrast gives the mass excess. Note the the time-dependence of $(aH)^2$ is exactly cancelled by the time-dependence of $\delta(r,t)$ - such that both $\delta_R$ and $\mathcal{C}(r)$ are time-independent qualities as required.


It can be argued that the sub-horizon damping of modes provides a natural smoothing to remove smaller modes, and that the height of peaks at horizon crossing can be used. However, this again involves applying a linear transfer function during horizon crossing, and the critical value should be determined at horizon crossing rather than the often used Musco criterion. Therefore, the answer to the question posed in the section title is that the density should be smoothed in order to calculate PBH abundance.

\section{Uncertainty in abundance due to the choice of smoothing function}
\label{sec:smoothingfunction}

In section \ref{sec:smooth}, it was argued that the use of a smoothing function to determine PBH formation is desirable in order to study a range of scales. However, it has been known for a long time that the specific choice of smoothing function has an effect on the calculated abundance of PBHs, as was investigated quantitatively in reference \cite{Ando:2018qdb}, and the effect of a window function in the context of non-Gaussianity is discussed in \cite{Atal:2018neu}. In that paper, it was found that the choice of smoothing function can mean that the calculated amplitude of the power spectrum required to produce the same abundance of PBHs can vary by more than an order of magnitude. Their calculation will be reproduced here, correcting for the fact that the critical value for collapse should be calculated with the same smoothing function that the variance is calculated with - important such that the numerator and denominator in equation \eqref{eqn:nu} are calculated in the same fashion. As a part of this calculation, corrected values for the formation criteria when different smoothing functions are used will also be given for the first time.

To obtain the the smoothed density contrast, the smoothing function is convolved with the density contrast to give the volume-averaged density contrast:
\begin{equation}
\delta_R = \int\mathrm{d}^3 x W(\vec{x},R)\delta(\vec{X}-\vec{x}) = \int\limits_0^{\infty}\mathrm{d}r 4 \pi r^2 W(r,R)\delta(r),
\label{eqn:deltasmooth}
\end{equation}
where the first and second equalities is in cartesian and radial coordinates respectively. For the second equality, it is assumed that a single spherically symmetric peak is considered, with the smoothing function centred on the peak (i.e. that the smoothing function is still being convolved with the density, but the spatial coordinate for the smoothed density is $r=0$ and as such does not appear explicitly in the equation).

The smoothing functions will be defined so that their heights and widths follow two rules. Firstly, where possible, the height of the smoothing function should be normalised such that the volume is equal to unity. Secondly, the width of each smoothing function is defined such that the $4 \pi r^2 W(r,R)$ component of the above integral should peak at a scale $R$ \footnote{Note that this means that the window functions described here differ slightly from those often used in the literature. For example, the standard Gaussian window function has an extra factor of $1/ \sqrt{2}$ inside the exponential compared to equation \eqref{eqn:gaussianWindow}. This results in the largest contribution to the integral in equation \eqref{eqn:deltasmooth} coming from the negative $\delta$ region which typically surrounds overdensities (see figure \ref{fig:shapes}) when smoothed on the scale $R=r_m$.}. 

\begin{figure*}
 \centering
  \includegraphics[width=\textwidth]{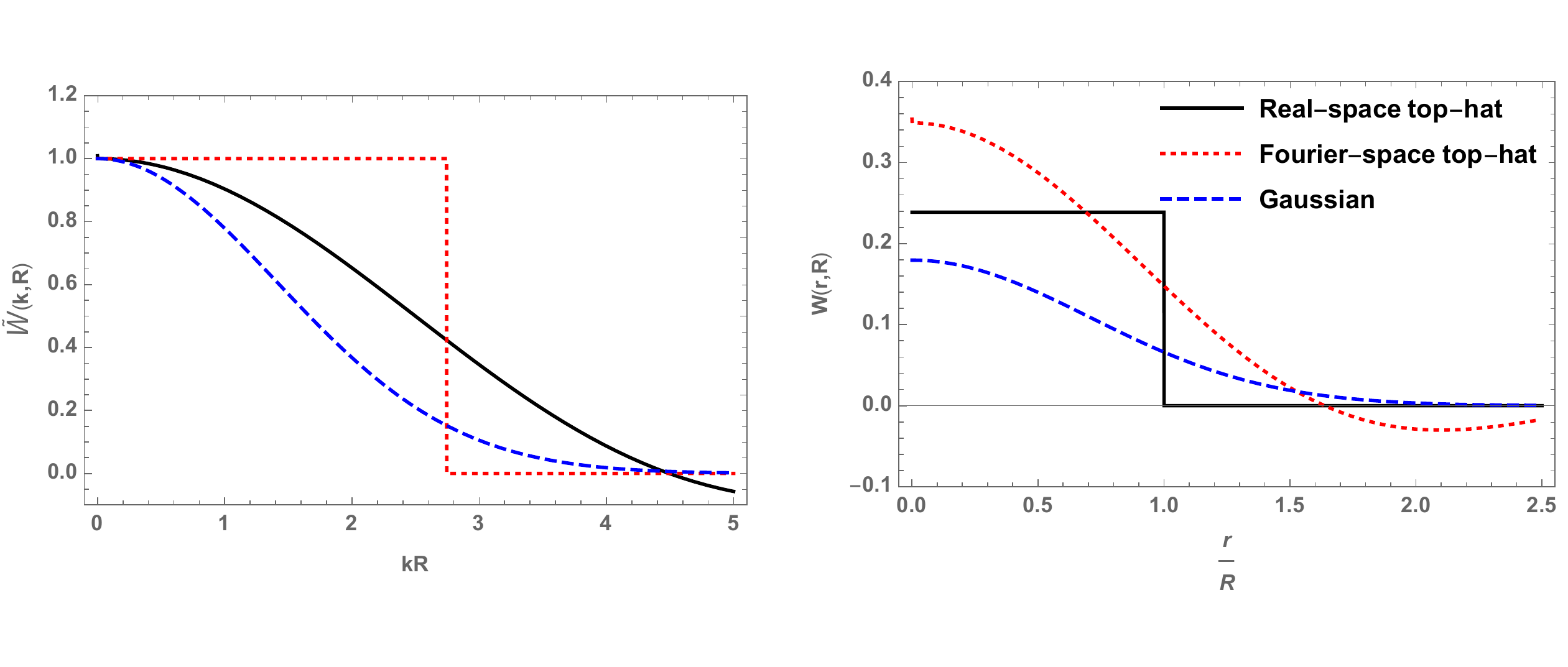} 
  \caption{ Different smoothing functions are plotted in Fourier-space (left) and real-space (right). The smoothing functions considered are: a real-space top-hat (black, solid); a Fourier-space top-hat (red, dotted); and a Gaussian (blue, dashed). The functions described here differ from those commonly used in the literature in that they have been normalised such the integrand $4 \pi r^2 W(r,R)$ in equation \eqref{eqn:deltasmooth} peaks at the smoothing scale $R$. }
  \label{fig:windows}
\end{figure*}

The choices of smoothing functions $W(r,R)$ and their Fourier transforms $\tilde{W}(k,R)$ (where $R$ represents the smoothing scale) are:
\begin{itemize}
\item \emph{real-space top-hat} (equivalent to the volume-averaging used for the Musco criterion):
\begin{equation}
W(r,R) = \frac{3}{4 \pi R^3}\Theta(R-r),
\end{equation}
\begin{equation}
\tilde{W}(k,R) = 3\frac{\mathrm{sin}(kR)-kR\mathrm{cos}(kR)}{(kR)^3},
\end{equation}

\item \emph{Fourier-space top hat},
\begin{equation}
W(r,R) = \frac{1}{2\pi^2 (\frac{R}{2.744})^3}\frac{\mathrm{sin}(r (\frac{R}{2.744})^{-1})-r (\frac{R}{2.744})^{-1}\mathrm{cos}(r (\frac{R}{2.744})^{-1})}{(r (\frac{R}{2.744})^{-1})^3},
\end{equation}
\begin{equation}
\tilde{W}(k,R) = \Theta((\frac{R}{2.744})^{-1}-k),
\end{equation}
where it is noted that the physical volume is unbound, and \cite{Ando:2018qdb} has been followed such that the normalisation $\tilde{W}(0,R)=1$ has been applied instead,

\item \emph{Gaussian}
\begin{equation}
W(r,R) = \frac{1}{(\pi R^2)^{3/2}}\exp\left( -\frac{r^2}{R^2} \right),
\label{eqn:gaussianWindow}
\end{equation}
\begin{equation}
\tilde{W}(k,R) = \exp \left( -\frac{(kR)^2}{4} \right).
\end{equation}

\end{itemize}
The left and right plots of figure \ref{fig:windows} shows the Fourier- and real-space smoothing functions respectively.

The most important factor contributing to the abundance of PBHs is $\nu_c$, the ratio of the variance of perturbations to the critical value. As described above, a full calculation of the abundance will not be considered here, but instead, for simplicity, the parameter $\nu$ will be compared for different choices of smoothing functions. First, lets consider the variance which would be calculated from using different smoothing variations. In this paper, only the simple case of a scale invariant curvature perturbation power spectrum, $\mathcal{P}_\mathcal{R}=\mathcal{A}_s$, will be considered, such that the smoothed density power spectrum during radiation domination is
\begin{equation}
\mathcal{P}_\delta=\frac{16}{81}\epsilon^4(t)(k r_m)^4\tilde{W}^2(k,r_m) T^2(k,\eta)\mathcal{A}_s,
\end{equation}
where the linear transfer function $T(k,\eta)$ has now been included to include the time-dependence of the power spectrum, $\eta$ is the conformal time. The transfer function on sub-horizon scales is given by \cite{Josan:2009qn}
\begin{equation}
T(k,\eta)=3\frac{\mathrm{sin}(k\eta/\sqrt{3})-(k\eta/\sqrt{3})\mathrm{cos}((k\eta/\sqrt{3}))}{((k\eta/\sqrt{3}))^3},
\end{equation}
and on super-horizon scales, $T(k,\eta)=1$. The variance of the smoothed density contrast is then given by integrating the power spectrum as in equation \eqref{eqn:powerintegral}. 

\begin{figure*}
 \centering
  \includegraphics[width=\textwidth]{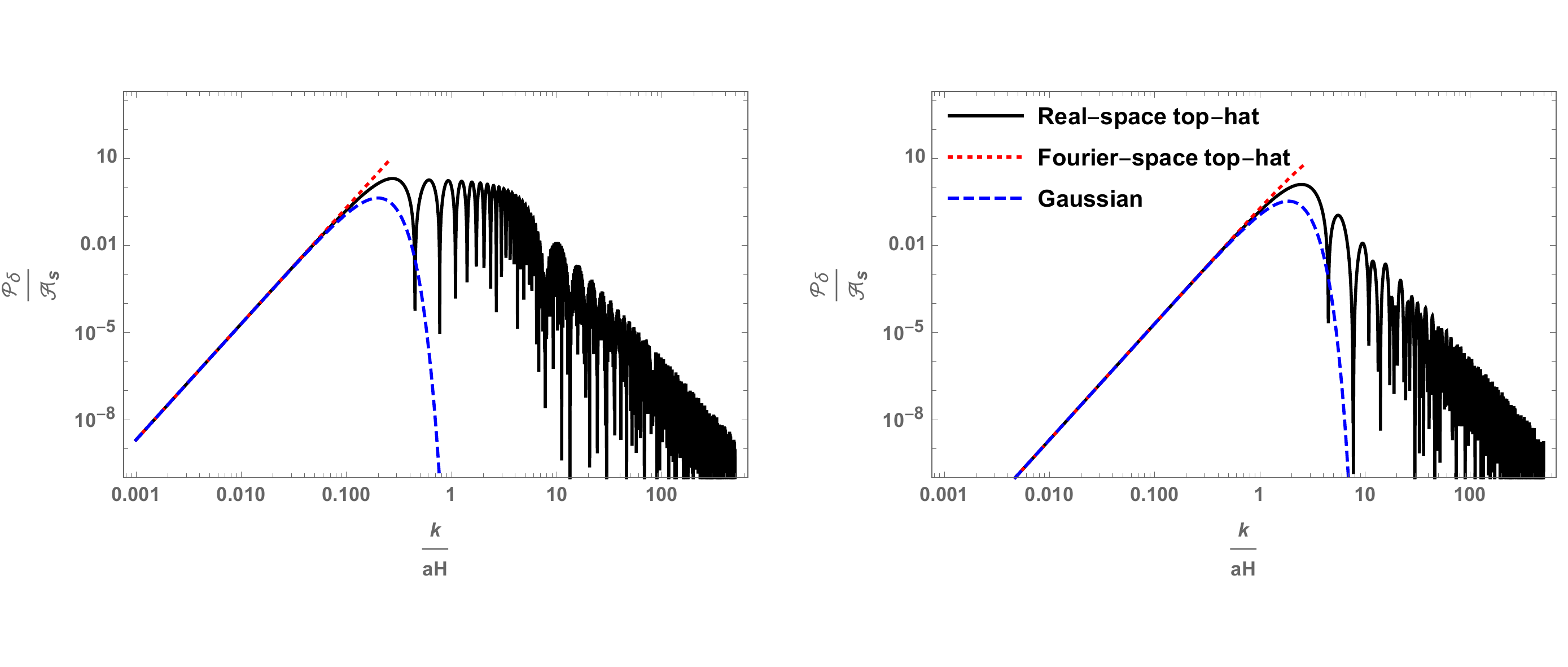} 
  \caption{ The effect of different smoothing functions applied to an initially scale-invariant power spectrum, applied on a super-horizon scale 10 times greater than the horizon (left) and on the horizon scale (right). In the left plot, the power spectrum has been multiplied by a factor $\epsilon^4=10 000$ to extract the time-independent component of the power-spectrum (as much as is possible). The smoothing functions considered are: a real-space top-hat (black, solid); a Fourier-space top-hat (red, dotted); and a Gaussian (blue, dashed).}
  \label{fig:integrand}
\end{figure*}

The smoothed power spectrum is plotted in figure \ref{fig:integrand}. Due to the presence of the transfer function, it is no longer possible to completely separate the time-dependent and time-independent components - but at least the factor of $\epsilon^4$ will be removed from the calculation. The left plot shows the integrand evaluated when the perturbation (and smoothing scale) are 10 times greater than the cosmological horizon - in the super-horizon regime as desired. 

We see that both the Fourier-space top-hat and Gaussian smoothing functions quickly reduce the amplitude of the power spectrum to zero - meaning that the transfer function has a negligible effect. However, for the real-space top-hat smoothing function, the power spectrum oscillates and does not converge to zero until sub-horizon modes are considered and the transfer function becomes important - in this scenario, modes which are significantly smaller than the perturbation (but greater than the horizon at the time of evaluation) erroneously contribute. Indeed, if one considers only the super-horizon limit (i.e. by taking the transfer function equal to unity), the integral to calculate the variance does not converge. For this reason, a Gaussian smoothing function has often been favoured in the literature (i.e. \cite{Young:2014ana,Green:2004wb}). When using a real-space top-hat smoothing function, the standard approach followed in the literature is to instead evaluate the variance at horizon crossing - meaning the perturbation scale, smoothing scale and horizon scale coincide\footnote{Though it is noted that this is typically not self-consistent, as the threshold value for collapse is normally stated in terms of the time-independent component in the super-horizon limit, and the use of a linear transfer for such large perturbations may not be valid at horizon crossing.}. This is shown in the right hand plot of figure \ref{fig:integrand}. For the Gaussian and Fourier-space top-hat smoothing functions, smoothing on the horizon scale instead of in the super-horizon limit has an $\mathcal{O}(10\%)$ effect on the calculated variance, due to the extra damping of sub-horizon modes by the transfer function.

If the variance is calculated using a given smoothing function, then the density should also be smoothed using the same smoothing function so that like parameters are being compared\footnote{The standard approach so far has been to use the same critical value $\delta_c$ (typically the Musco criterion) regardless of which smoothing function is used - which led to the analysis in \cite{Ando:2018qdb}.}. In this paper, the different critical values are obtained by applying the smoothing functions to the density contrast, as in equation \eqref{eqn:deltasmooth}. The sinc function profile given in equation \eqref{eqn:sinc} will be used, and the value obtained in \cite{Germani:2018jgr} for a perturbation from a narrow-peak power spectrum will be used, giving a critical amplitude $A\approx 1.2$ in order to collapse. 

Table \ref{tab} shows a comparison of the values obtained from using different smoothing functions - the variance of perturbations if evaluated at horizon crossing (noted by HC) or in the super-horizon limit (noted by SH), the critical density for collapse $\delta_c$, and finally the relative peak height for collapse $\nu_c$. Where necessary, the values are scaled to the amplitude of the scale invariant power spectrum, $\mathcal{A}_s$. Ideally, all of the calculations would agree - and we would see little to no variation in $\nu_c$. 

Reference \cite{Ando:2018qdb} found that, for a fixed $\mathcal{A}_s$, $\nu_c$ varies by around an order of magnitude when different window functions are used - and this result is confirmed when a constant $\delta_c$ is used. However, when accounting for the fact that $\delta_c$ should be calculated with the same function, we see that $\nu_c$ changes only by a factor $\mathcal{O}(10\%)$. This means that the uncertainty in $\mathcal{A}_s$ required to produce a given abundance of PBHs due to the choice of window function is also $\mathcal{O}(10\%)$, which is of the same order as the uncertainty due to the critical value \cite{Musco:2018rwt}. Note that this analysis does not make an argument for which is the correct window function to use, but only quantifies the uncertainty involved, such an investigation is left for future study.

\begin{table}[t]
\begin{center}
\begin{tabular}{p{4.5cm}|p{2cm} p{2cm} p{2cm}}

\textbf{Smoothing function} & $\mathbf{\langle\delta_R^2\rangle}/\mathcal{A}_s$ & $\mathbf{\delta_c}$ & $\mathbf{\mathcal{A}_s^{1/2}\nu_c}$  \\ \midrule
\hline
Real-space top-hat (HC) & 1.06 & 0.51 & 0.49  \\ \midrule
Fourier-space top-hat (HC) & 2.00 & 0.59 & 0.41  \\ \midrule
Fourier-space top-hat (SH) & 2.80 & 0.59 & 0.35  \\ \midrule
Gaussian (HC) & 0.31 & 0.18 & 0.33  \\ \midrule
Gaussian (SH) & 0.40 & 0.18 & 0.29  \\ \bottomrule

\end{tabular}
\end{center}
\caption{The use of different smoothing functions at different times is considered. The power spectrum is taken to be scale-invariant with an amplitude $\mathcal{A}_s$. The variance of the ``time-independent'' density contrast is evaluated in the super-horizon limit (SH) and at horizon crossing (HC). This is compared to the amplitude of the threshold value $\delta_c$ calculated with each smoothing function. The relative height $\nu_c$ paramterises the heights of peaks relative to the variance required to form a PBH.}
\label{tab}
\end{table}

\section{Formation criterion from the curvature perturbation}
\label{sec:zeta}

The formation criterion in terms of the metric perturbation $\mathcal{R}$ will now be discussed. In the comoving uniform-density slicing, the metric appears as \footnote{Here $\mathcal{R}$ is used to denote the curvature perturbation, whilst $\zeta$ is often used on uniform-density hypersurfaces. On super-horizon scales, the density contrast goes to zero, and it can be considered that $\mathcal{R}=\zeta$, and $\mathcal{R}$ and $\zeta$ are often used interchangeably in the literature.}
\begin{equation}
\mathrm{d}s^2 = -\mathrm{d}t^2 + a^2(t) \exp (2\zeta) \left( \mathrm{d}\hat{r}^2+\hat{r}^2\mathrm{d}\Omega^2 \right),
\end{equation}
where $\Omega$ contains the angular dependence, and $\hat{r}$ is the radial coordinate. Note that the radial coordinate $\hat{r}$ is related to the comoving radial coordinate $r$ as
\begin{equation}
r = \exp(\mathcal{R}(\hat{r}))\hat{r}.
\end{equation}
On super-horizon scales, the density contrast during radiation domination is then given by \cite{Musco:2018rwt}
\begin{equation}
\delta(r) = -\frac{4}{9}\left( \frac{1}{aH} \right)^2\exp(-2\mathcal{R}(\hat{r}))\left( \mathcal{R}''(\hat{r})+\frac{2\mathcal{R}'(\hat{r})}{\hat{r}}+\frac{1}{2}\mathcal{R}'^2(\hat{r}) \right),
\label{eqn:nonlinear}
\end{equation}
where the prime denotes a derivative with respect to $\hat{r}$ and spherical symmetry has been assumed.

It can be seen that $\delta(r)$ appears to depend upon the absolute value of $\mathcal{R}$ inside the exponential. Naively, this leads to the conclusion that if a constant positive value is added to (subtracted from) $\mathcal{R}$ we would see significantly smaller (larger) density perturbations, in turn leading to significantly more (less) PBHs forming. However, according to the separate universe approach \cite{Rigopoulos:2003ak}, adding (or subtracting) a constant value to $\mathcal{R}$ should instead only correspond to a time-shift - rescaling the background energy density and horizon scale. This can be seen if $\epsilon$ is substituted into the expression,
\begin{equation}
\epsilon(t) = \frac{1}{aH \exp(\mathcal{R}(\hat{r}_m))\hat{r}_m},
\end{equation}
which gives
\begin{equation}
\delta(r) = -\frac{4}{9}\epsilon(t)^2\exp(2(\mathcal{R}(\hat{r}_m)-\mathcal{R}(\hat{r})))\hat{r}_m^2\left( \mathcal{R}''(\hat{r})+\frac{2\mathcal{R}(\hat{r})}{\hat{r}}+\frac{1}{2}\mathcal{R}'^2(\hat{r}) \right).
\end{equation}
The term $(\mathcal{R}(\hat{r}_m)-\mathcal{R}(\hat{r}))$ now appears inside the exponential - indicating that PBH formation is not sensitive to constant values added to $\mathcal{R}$, and further that $\mathcal{R}$-modes significantly larger than $\hat{r}_m$ do not affect PBH formation either. Indeed, one can consider a perturbation in the super-horizon limit and consider a small patch at the centre of a perturbation. In this region, $\mathcal{R}$ can be considered constant and the region thought of a separate universe, with a horizon scale given by $r_H=(aH)^{-1}\exp(-\mathcal{R})$. When considering peaks in the super-horizon limit then, the exponential term can be considered simply as a rescaling of the horizon scale (such that the peaks in regions with positive (negative) $\mathcal{R}$ have a smaller (larger) initial amplitude, but the horizon is smaller (larger) and thus the perturbation grows more (less) in the longer (shorter) time before the perturbation reenters the horizon). Considering the height of peaks of $\delta$ at a specific early time (while a perturbation is still super-horizon) therefore may not give a suitable indication of whether such a peak will eventually collapse to form a PBH or not.

We will now consider the effect of smoothing the distribution. This is most easily done for the case of the real-space top-hat window function, where it is now important to make the distinction that the volume-averaging is completed by integrating over the comoving coordinate $r$ \footnote{Using the Areal radius $R=a(t)r=a(t)\exp(\mathcal{R}(\hat{r}))\hat{r}$ is correct, but the scale factor divides out when divided by the volume, giving no difference.}. It can be shown that the top-hat smoothing gives \cite{Musco:2018rwt}
\begin{equation}
\delta_R = \frac{1}{V}\int\limits_0^{r_m}\mathrm{d}r 4 \pi r^2 \delta(r,t) = \epsilon^2(t) \left( \left(-\frac{2}{3}\hat{r}_m\mathcal{R}'(\hat{r}_m)\right)-\frac{3}{8}\left( -\frac{2}{3}\hat{r}_m\mathcal{R}'(\hat{r}_m) \right)^2 \right).
\label{eqn:densitycurvature}
\end{equation}
Due to the integration over the radial coordinate $r$ (and not $\hat{r}$), the exponential term factors out of the final expression. By removing the factor of $\epsilon^2(t)$, the time-independent component can be extracted. This expression represents a quadratic relationship between $\delta$ and $\mathcal{R}$, where $\left(-\frac{2}{3}\hat{r}_m\mathcal{R}'(\hat{r}_m)\right)$ is equal to the standard expression derived in the linear case. The effect of the non-linear term suppresses PBH formation and requires that the power spectrum have a larger amplitude by an approximate factor $f$ compared to the linear case \cite{Young:2019yug}, where $f$ is given by
\begin{equation}
f = \left( \frac{4\left(1 - \sqrt{\frac{2-3\delta_c}{2}} \right)}{3\delta_c} \right)^2,
\end{equation}
where $f\approx1.81$ for the value $\delta_c\approx0.51$ used in this paper. It is noted here that equation \eqref{eqn:nonlinear} is valid in the linear regime, whilst the abundance of PBHs is typically evaluated at horizon crossing, in order to ensure that the variance converges (see section \ref{sec:smoothingfunction}).

For other window functions, the integral is more complicated, and typically it is not possible to solve analytically for a general profile shape. The consideration of the non-Gaussianity arising from the non-linear terms when other smoothing functions are considered goes beyond the scope of this paper. However, the non-Gaussianity is not expected to be significantly different.

\section{Summary}
\label{sec:summary}
The key conclusions of the topics discussed in this paper are summarised below:

\begin{enumerate}

\item \emph{The average profile shape of primordial perturbations}: the central region of primordial perturbations can be estimated from the power spectrum $\mathcal{P}_\delta$. For the power spectra considered here (representative of those considered in the literature), the central regions of the resultant profiles all share a similar shape well represented (in the central region) by a sinc function, equation \eqref{eqn:sinc}. This similarity is due to similarities in the power spectrum considered - which are all well described by a narrow peak in $k$-space. See figure \ref{fig:shapes} for a comparison of the power spectrum and profile shapes. The shape of perturbations arising from modes in the centre of a broad peaked power spectrum remains unknown because the calculated average profile shape is dominated by the smallest-scale modes considered due to the $k^4$ growth of $\mathcal{P}_\delta$.

\item \emph{At what time should the abundance be calculated?} Papers calculating the threshold amplitude for collapse $\delta_c$ typically state this value in terms of the amplitude of the initial perturbations in the super-horizon limit, using the time-independent component of the density contrast. In addition, as they near horizon crossing, perturbations large enough to form PBHs display complicated behaviour, and the use of a critical value calculated in the linear regime and a linear transfer function may not be valid. For these reasons, it is desirable to calculate the abundance of PBHs that will form in the super-horizon limit, during the linear regime.

\item \emph{PBH formation time:} PBHs forming from perturbations of identical scales do not form at exactly the same time, due to variance in the formation time and the time of horizon crossing of the perturbations. For this reason, it is not possible to state an exact time at which PBHs form on a given scale. Here, it has been argued that, for simplicity, the horizon crossing time should be taken as the formation time (rather than the somewhat later time at which an event horizon forms) - allowing easier comparison of initial PBH abundance and derivation of constraints.

\item \emph{Volume-averaged (smoothed) or peak value:} the use of both parameters is discussed and it is concluded that using the smoothed density is preferable because it allows for PBH formation at a range of times and scales to be calculated with the tools currently available. To make the same calculation using the central peak value would require knowledge of the full non-linear transfer function at horizon crossing, the critical amplitude of a peak at horizon crossing, and the non-linear evolution of the horizon scale. It is also noted that the critical amplitude of the smoothed density contrast shows less variation when different profile shapes are considered, lying in the range $0.41<\delta_c<2/3$. On the other hand, it can only be stated that the critical peak amplitude $\delta_{c,pk}>2/3$ for centrally peaked perturbations (and in principle, for off-centred peaks, it is also seen that smaller values are found \cite{Young:2019yug}).

\item \emph{The uncertainty due to the choice of smoothing function.} It is important that the smoothing function used when determining the variance of perturbations should be the same as the smoothing function used when calculating the formation criterion. Table \ref{tab} shows the formation criteria which should be used for different choices smoothing functions, and compares this to the variance calculated for a scale-invariant power spectrum $\mathcal{P}_\mathcal{R}=\mathcal{A}_s$. In this paper, the use of real-space top-hat, Fourier-space top-hat and Gaussian smoothing functions has been compared, finding that this results in an uncertainty of $\mathcal{O}(10\%)$ in the amplitude of the power spectrum. 


\item \emph{The formation criterion in terms of the curvature perturbation $\mathcal{R}$.} The curvature perturbation can be related to the unsmoothed and smoothed density contrast using equations \eqref{eqn:nonlinear} and \eqref{eqn:densitycurvature} respectively. Again, it is concluded that the the smoothed time-independent component of the density contrast $\delta_{R,TI}$ gives a more robust criterion for determining whether a PBH will form, which is related to the curvature perturbation as
\begin{equation}
\delta_{R,TI} = \delta_1-\frac{3}{8}\delta_1^2,
\end{equation}
with $\delta_1$ representing the linear component,
\begin{equation}
\delta_1=-\frac{2}{3}\hat{r}_m\mathcal{R}'(\hat{r}_m)
\end{equation}
which is valid for spherically symmetric perturbations, and a real-space top-hat smoothing function has been used.

\end{enumerate}

\section*{Acknowledgements}
Christian T. Byrnes, Ilia Musco, and Nicola Bellomo are thanked for helpful discussions. Eiichiro Komatsu and Christian T. Byrnes are thanked for their comments on a draft of this paper. SY is supported by a Humboldt Research Fellowship for Postdoctoral Researchers.


\bibliographystyle{ieeetr} 
\bibliography{bibfile}

\begin{thebibliography}{10}

\bibitem{Carr:1974nx}
B.~J. Carr and S.~W. Hawking, ``{Black holes in the early Universe},'' {\em
  Mon. Not. Roy. Astron. Soc.}, vol.~168, pp.~399--415, 1974.

\bibitem{Musco:2004ak}
I.~Musco, J.~C. Miller, and L.~Rezzolla, ``{Computations of primordial black
  hole formation},'' {\em Class. Quant. Grav.}, vol.~22, pp.~1405--1424, 2005.

\bibitem{Musco:2008hv}
I.~Musco, J.~C. Miller, and A.~G. Polnarev, ``{Primordial black hole formation
  in the radiative era: Investigation of the critical nature of the
  collapse},'' {\em Class. Quant. Grav.}, vol.~26, p.~235001, 2009.

\bibitem{Musco:2012au}
I.~Musco and J.~C. Miller, ``{Primordial black hole formation in the early
  universe: critical behaviour and self-similarity},'' {\em Class. Quant.
  Grav.}, vol.~30, p.~145009, 2013.

\bibitem{Musco:2018rwt}
I.~Musco, ``{The threshold for primordial black holes: dependence on the shape
  of the cosmological perturbations},'' 2018.

\bibitem{Harada:2015ewt}
T.~Harada and S.~Jhingan, ``{Spherical and nonspherical models of primordial
  black hole formation: exact solutions},'' {\em PTEP}, vol.~2016, no.~9,
  p.~093E04, 2016.

\bibitem{Harada:2015yda}
T.~Harada, C.-M. Yoo, T.~Nakama, and Y.~Koga, ``{Cosmological long-wavelength
  solutions and primordial black hole formation},'' {\em Phys. Rev.}, vol.~D91,
  no.~8, p.~084057, 2015.

\bibitem{Nakama:2013ica}
T.~Nakama, T.~Harada, A.~G. Polnarev, and J.~Yokoyama, ``{Identifying the most
  crucial parameters of the initial curvature profile for primordial black hole
  formation},'' {\em JCAP}, vol.~1401, p.~037, 2014.

\bibitem{Nakama:2014fra}
T.~Nakama, ``{The double formation of primordial black holes},'' {\em JCAP},
  vol.~1410, no.~10, p.~040, 2014.

\bibitem{Shibata:1999zs}
M.~Shibata and M.~Sasaki, ``{Black hole formation in the Friedmann universe:
  Formulation and computation in numerical relativity},'' {\em Phys. Rev.},
  vol.~D60, p.~084002, 1999.

\bibitem{Niemeyer:1999ak}
J.~C. Niemeyer and K.~Jedamzik, ``{Dynamics of primordial black hole
  formation},'' {\em Phys. Rev.}, vol.~D59, p.~124013, 1999.

\bibitem{Polnarev:2006aa}
A.~G. Polnarev and I.~Musco, ``{Curvature profiles as initial conditions for
  primordial black hole formation},'' {\em Class. Quant. Grav.}, vol.~24,
  pp.~1405--1432, 2007.

\bibitem{Green:2004wb}
A.~M. Green, A.~R. Liddle, K.~A. Malik, and M.~Sasaki, ``{A New calculation of
  the mass fraction of primordial black holes},'' {\em Phys. Rev.}, vol.~D70,
  p.~041502, 2004.

\bibitem{Young:2014ana}
S.~Young, C.~T. Byrnes, and M.~Sasaki, ``{Calculating the mass fraction of
  primordial black holes},'' {\em JCAP}, vol.~1407, p.~045, 2014.

\bibitem{Harada:2013epa}
T.~Harada, C.-M. Yoo, and K.~Kohri, ``{Threshold of primordial black hole
  formation},'' {\em Phys. Rev.}, vol.~D88, no.~8, p.~084051, 2013.
\newblock [Erratum: Phys. Rev.D89,no.2,029903(2014)].

\bibitem{Bardeen:1985tr}
J.~M. Bardeen, J.~R. Bond, N.~Kaiser, and A.~S. Szalay, ``{The Statistics of
  Peaks of Gaussian Random Fields},'' {\em Astrophys. J.}, vol.~304,
  pp.~15--61, 1986.

\bibitem{Byrnes:2018clq}
C.~T. Byrnes, M.~Hindmarsh, S.~Young, and M.~R.~S. Hawkins, ``{Primordial black
  holes with an accurate QCD equation of state},'' {\em JCAP}, vol.~1808,
  no.~08, p.~041, 2018.

\bibitem{Germani:2018jgr}
C.~Germani and I.~Musco, ``{The abundance of primordial black holes depends on
  the shape of the inflationary power spectrum},'' {\em Phys. Rev. Lett.},
  vol.~122, no.~14, p.~141302, 2019.

\bibitem{Byrnes:2018txb}
C.~T. Byrnes, P.~S. Cole, and S.~P. Patil, ``{Steepest growth of the power
  spectrum and primordial black holes},'' 2018.

\bibitem{Young:2013oia}
S.~Young and C.~T. Byrnes, ``{Primordial black holes in non-Gaussian
  regimes},'' {\em JCAP}, vol.~1308, p.~052, 2013.

\bibitem{Young:2014oea}
S.~Young and C.~T. Byrnes, ``{Long-short wavelength mode coupling tightens
  primordial black hole constraints},'' {\em Phys. Rev.}, vol.~D91, no.~8,
  p.~083521, 2015.

\bibitem{Young:2019yug}
S.~Young, I.~Musco, and C.~T. Byrnes, ``{Primordial black hole formation and
  abundance: contribution from the non-linear relation between the density and
  curvature perturbation},'' 2019.

\bibitem{Yoo:2018esr}
C.-M. Yoo, T.~Harada, J.~Garriga, and K.~Kohri, ``{Primordial black hole
  abundance from random Gaussian curvature perturbations and a local density
  threshold},'' {\em PTEP}, vol.~2018, no.~12, p.~123, 2018.

\bibitem{Kawasaki:2019mbl}
M.~Kawasaki and H.~Nakatsuka, ``{Effect of nonlinearity between density and
  curvature perturbations on the primordial black hole formation},'' 2019.

\bibitem{DeLuca:2019qsy}
V.~De~Luca, G.~Franciolini, A.~Kehagias, M.~Peloso, A.~Riotto, and C.~Ünal,
  ``{The Ineludible non-Gaussianity of the Primordial Black Hole Abundance},''
  2019.

\bibitem{Lyth:2004gb}
D.~H. Lyth, K.~A. Malik, and M.~Sasaki, ``{A General proof of the conservation
  of the curvature perturbation},'' {\em JCAP}, vol.~0505, p.~004, 2005.

\bibitem{Nakama:2016gzw}
T.~Nakama, J.~Silk, and M.~Kamionkowski, ``{Stochastic gravitational waves
  associated with the formation of primordial black holes},'' {\em Phys. Rev.},
  vol.~D95, no.~4, p.~043511, 2017.

\bibitem{Ando:2018qdb}
K.~Ando, K.~Inomata, and M.~Kawasaki, ``{Primordial black holes and
  uncertainties in the choice of the window function},'' {\em Phys. Rev.},
  vol.~D97, no.~10, p.~103528, 2018.

\bibitem{Atal:2018neu}
V.~Atal and C.~Germani, ``{The role of non-gaussianities in Primordial Black
  Hole formation},'' {\em Phys. Dark Univ.}, p.~100275, 2018.

\bibitem{Josan:2009qn}
A.~S. Josan, A.~M. Green, and K.~A. Malik, ``{Generalised constraints on the
  curvature perturbation from primordial black holes},'' {\em Phys. Rev.},
  vol.~D79, p.~103520, 2009.

\bibitem{Rigopoulos:2003ak}
G.~I. Rigopoulos and E.~P.~S. Shellard, ``{The separate universe approach and
  the evolution of nonlinear superhorizon cosmological perturbations},'' {\em
  Phys. Rev.}, vol.~D68, p.~123518, 2003.

\end{thebibliography}

\end{document}